\documentclass[aps,prb,twocolumn,doublespace,showpacs,superscriptaddress,groupedaddress,floatfix]{revtex4-1}
\usepackage{graphicx}
\usepackage{color}
\usepackage{amsmath}
\usepackage{amssymb}
\usepackage{comment}
\usepackage{gensymb}
\usepackage{longtable}
\usepackage{setspace}
\usepackage{tikz}
\usetikzlibrary{shapes,arrows}
\usepackage{titlesec}

\definecolor{scarred}{rgb}{0.75,0.0,0.0}
\begin{document}
\title{A first principles investigation of cubic BaRuO$_3$: A Hund's metal}
\author{Nagamalleswararao Dasari$^{1}$}\email{nagamalleswararao.d@gmail.com}
\author{S.\ R.\ K.\ C.\ Sharma Yamijala$^{2}$}
\author{Manish Jain$^{3}$}
\author{T. Saha Dasgupta$^{4}$}
\author{Juana Moreno$^{5,6}$}
\author{Mark Jarrell$^{5,6}$}
\author{N.\ S.\ Vidhyadhiraja$^{1}$}\email{raja@jncasr.ac.in}
\affiliation{$^{1}$Theoretical Sciences Unit, Jawaharlal Nehru Centre For
Advanced Scientific Research, Jakkur, Bangalore 560064, India.}
\affiliation{$^{2}$ Chemistry and Physics of Materials Unit, Jawaharlal Nehru Centre For Advanced Scientific Research, Jakkur, Bangalore 560064, India.}
\affiliation{$^{3}$ Department of Physics, Indian Institute of Science, Bangalore 560012, India}
\affiliation{$^{4}$ S.\ N.\ Bose Centre for Basic Sciences, Kolkata 700 098, India.}
\affiliation{$^{5}$ Department of Physics $\&$ Astronomy, Louisiana State University, Baton Rouge, LA 70803-4001, USA.}
\affiliation{$^{6}$ Center for Computation and Technology, Louisiana State University, Baton Rouge, LA 70803, USA.}

\begin{abstract}

A first-principles investigation of cubic-BaRuO$_3$, by 
combining density functional theory with dynamical mean-field theory 
and a hybridization expansion continuous time quantum Monte-Carlo solver, 
has been carried out. Non-magnetic calculations with appropriately 
chosen on-site Coulomb repulsion, $U$ and Hund's exchange, $J$, for 
single-particle dynamics and static susceptibility show that
cubic-BaRuO$_3$ is in a spin-frozen state at temperatures 
above the ferromagnetic transition point. A strong red shift with 
increasing $J$ of the peak in the real frequency dynamical
 susceptibility indicates a dramatic suppression of the Fermi 
liquid coherence scale as compared to the bare parameters
 in cubic-BaRuO$_3$. The self-energy also shows clear 
deviation from Fermi liquid behaviour that manifests in 
the single-particle spectrum. Such a clean separation of energy 
scales in this system provides scope for an incoherent
 spin-frozen (SF) phase, that extends over a wide temperature range, to manifest in 
 non-Fermi liquid behaviour and to be the precursor for the 
 magnetically ordered ground state. 
\end{abstract}

\maketitle

\section{Introduction}

Transition metal oxides (TMOs) have occupied a unique and very 
significant position in the investigations of correlated 
electron systems. The interplay of spin, charge and orbital 
degrees of freedom in the partially filled and localized 
3d and 4d orbitals leads to a rich set of phenomena including 
high temperature superconductivity, colossal magneto-resistance 
and the Mott metal-insulator transition.
Due to the extended nature of 4d orbitals, the corresponding 
TMOs exhibit strong hybridization with oxygen.
This leads to a large crystal field splitting that could be of the order of the
 local screened Coulomb interaction($U$) and a broad 4d band of width $W$.
As a consequence, these materials prefer a low spin state rather than the high spin state. 

Furthermore, the wide d-band in 4d-orbital based TMOs such as 
Ruthenates leads to a moderate screened Coulomb interaction $U\simeq W$ as 
compared to the much narrower d-band in 3d-orbital based TMOs\cite{Georges}.
Surprisingly however, most of the Ru-based TMOs show strong correlation effects that are reflected
 in the enhanced linear coefficient of specific heat $\gamma$.
A few of such ruthenates are mentioned in table-\ref{tab:specific_heat}, where we have also indicated 
the magnetic order of the ground state as well as the effective mass computed as the ratio of 
experimentally\cite{Georges,Zhao20072816} measured 
$\gamma$ to $\gamma_{LDA}$, computed\cite{Zhao20072816} within a local density approximation(LDA). 
The origin of such enhanced effective mass could be a local Coulomb repulsion induced proximity 
to a insulating state. An alternative origin could be 
Hund's\cite{PhysRevLett.101.166405,1367-2630-11-2-025021,PhysRevLett.107.256401,Georges} 
coupling $J$(intra-atomic exchange), which, as has been shown recently for several materials, 
especially Ruthenates\cite{PhysRevLett.101.166405,PhysRevLett.106.096401,PhysRevB.91.195149}, leads to their characterization 
as `Hund's metals'. A prominent member of this class is BaRuO$_3$ which,  
\begin{table}[h!]
\begin{center}
\setlength{\tabcolsep}{7.5pt}
\renewcommand{\arraystretch}{1.5}
\caption{Magnetic ground state and the ratio of $\gamma$ to $\gamma_{LDA}$ for 4d Ru-based compounds} 
\begin{ruledtabular}
\begin{tabular}{|c|c|c|}
\hline
Compound & Magnetic order & $\frac{\gamma}{\gamma_{LDA}}$ \\ \hline
Sr$_2$RuO$_4$ & PM & 4 \\
Sr$_3$Ru$_2$O$_7$ & PM & 10\\
CaRuO$_3$ & PM & 7 \\
SrRuO$_3$ & FM $<$ 160 K & 4 \\
3C-BaRuO$_3$ & FM $<$ 60 K & -- \\
4H-BaRuO$_3$ & PM & 3.37 \\
6H-BaRuO$_3$ & PM & 3.37 \\
9R-BaRuO$_3$ & PI & 1.54 \\
\hline
\end{tabular}
\label{tab:specific_heat}
\end{ruledtabular}
\end{center}
\end{table}
depending on synthesis conditions, can exist in four polytypes\cite{Jin20052008}. 
These are nine-layered rhombohedral (9R), four-layered hexagonal(4H), six-layered 
hexagonal(6H) and cubic(3C). The 9R has a paramagnetic insulating (PI) ground 
state while 4H and 6H are paramagnetic metals(PM).

The 3C-BaRuO$_3$ polytype is a ferromagnetic metal with Curie temperature, T$_c$ = 60 K, which is
much smaller than the value of T$_c$(= 160 K) in SrRuO$_3$\cite{PhysRevB.56.321}. The experimental
value of the saturated magnetic moment of 3C-BaRuO$_3$\cite{Jin20052008} is 0.8 $\mu_B$/Ru, which is
far less than 2.8 $\mu_B$/Ru, expected for a low spin state of 4d Ru. It is also smaller than
measured value of 1.4 $\mu_B$/Ru in SrRuO$_3$\cite{PhysRevB.56.321}. The observed effective magnetic
moment ($\mu_{eff}$) in the paramagnetic phase of BaRuO$_3$ and SrRuO$_3$, is however, very close to
the S=1 moment.  From table \ref{tab:specific_heat}, we can readily understand that electron
correlations in 4H-BaRuO$_3$ and 6H-BaRuO$_3$ are comparable with SrRuO$_3$ and in case of
9R-BaRuO$_3$ they are weak. Although the strength of electron correlations in 3C-BaRuO$_3$ is still
unknown, a non-Fermi liquid behavior in the experimental measured
resistivity\cite{Jin20052008,PhysRevLett.101.077206}, i.e., $\rho$(T) $\propto$ T$^{1.85}$ in the
ferromagnetic phase and its cross-over to T$^{0.5}$ in the paramagnetic phase
(similar to SrRuO$_3$\cite{PhysRevB.53.4393} and CaRuO$_3$\cite{PhysRevB.66.041104} compounds),
hints towards a strongly correlated system\cite{Georges}.

In the present work, the following questions have been addressed: Is 3C-BaRuO$_3$ a 
correlated metal or not? If yes, then what is the origin and strength of 
correlations? What is the probable origin of non-Fermi liquid NFL) signature 
in the resistivity\cite{Jin20052008,PhysRevLett.101.077206}?
We have employed the dynamical mean 
field theory(DMFT) framework in combination with an {\em{ab initio}} method\cite{kotliar}, 
namely density functional theory(DFT) within the generalized gradient approximation 
(GGA)\cite{perdew}. In the DMFT\cite{RevModPhys.68.13} framework, a lattice problem 
may be mapped on to a single impurity Anderson model with a self-consistently determined 
bath. The resulting quantum impurity problem has been solved by 
using hybridization expansion\cite{Gull2,Comanac} 
continuous-time quantum Monte-Carlo algorithm (HY-CTQMC). 
The main finding is that 3C-BaRuO$_3$ is a Hund's 
correlated metal. Furthermore we find that 3C-BaRuO$_3$ 
is in a spin-frozen state at temperatures in the neighbourhood of the experimental 
ferromagnetic transition 
temperature. This state, we speculate, is the precursor of the ferromagnetic ground state and 
also a possible origin of the experimentally 
observed NFL behavior in resistivity.

The rest of the paper is organised as follows. In section II, we describe 
the DFT details and Wannier projection briefly. In Section III, we describe 
our results from GGA+DMFT(CTQMC) for 3C-BaRuO$_3$. We present our conclusions in the final section.

\section{Details of the density functional theory calculations and results}
The 3C polytype of BaRuO$_3$ belongs to the space group of $Pm$-$3m$ which corresponds 
to an  ideal cubic perovskite structure, while the closely related
CaRuO$_3$ and SrRuO$_3$ crystallize in an orthorhombic distorted perovskite structure of space 
group $Pnma$\cite{Jin20052008}. A significant structural change from
 CaRuO$_3$ to 
SrRuO$_3$ and to BaRuO$_3$ is a decrease in bending angle\cite{Jin20052008} (180$^{\circ}$-$\phi$) of Ru-O-Ru 
bonds, which becomes zero for BaRuO$_3$. Apart from slight distortions of RuO$_6$ octahedra 
in CaRuO$_3$ and SrRuO$_3$, that are absent in 
BaRuO$_3$\cite{Jin20052008}, 
each of these materials have threefold 
degenerate t$_{2g}$ bands near the Fermi-level with a formal valance of 4 electrons\cite{Jin20052008} 
i.e., t$^4_{2g}$e$^0_g$.
Density functional theory (DFT) calculations have been performed within 
the generalized gradient approximation using the plane wave pseudo-potential 
code QUANTUM ESPRESSO\cite{QE-2009}.
 We have used ultra-soft pseudo-potentials with 
Perdew-Burke-Ernzerhof\cite{PhysRevB.54.16533} exchange-correlation functional. 
An 8$\times$8$\times$8 Monkhorst-Pack 
k-grid is used for optimization together with an 80 Ry energy cutoff and a 640 Ry 
charge cutoff. The system is considered to be optimized if the forces acting on 
all the atoms are less than 10$^{-4}$ Ry/Bohr. After optimization, we find the lattice parameter 
to be 4.0745 $\AA$. Throughout the calculations, Marzari-Vanderbilt cold smearing is used with 
a degauss value of 0.01 Ry. A 20$\times$20$\times$20 k-grid without any symmetries is used for all the 
nonself-consistent calculations (including Wannier90 calculations).
To extract the information of the low-energy subspace,
which will be used by the DMFT code, we have projected
the Bloch wave-functions obtained from our DFT calculations on to 
the Ru-t$_{2g}$ orbitals using the maximally localized Wannier functions\cite{PhysRevB.56.12847} (MLWF) 
technique as implemented in the Wannier90 code\cite{Mostofi2008685}.
\begin{figure}[t]
\centering
\includegraphics[angle=0,width=1.0\columnwidth]{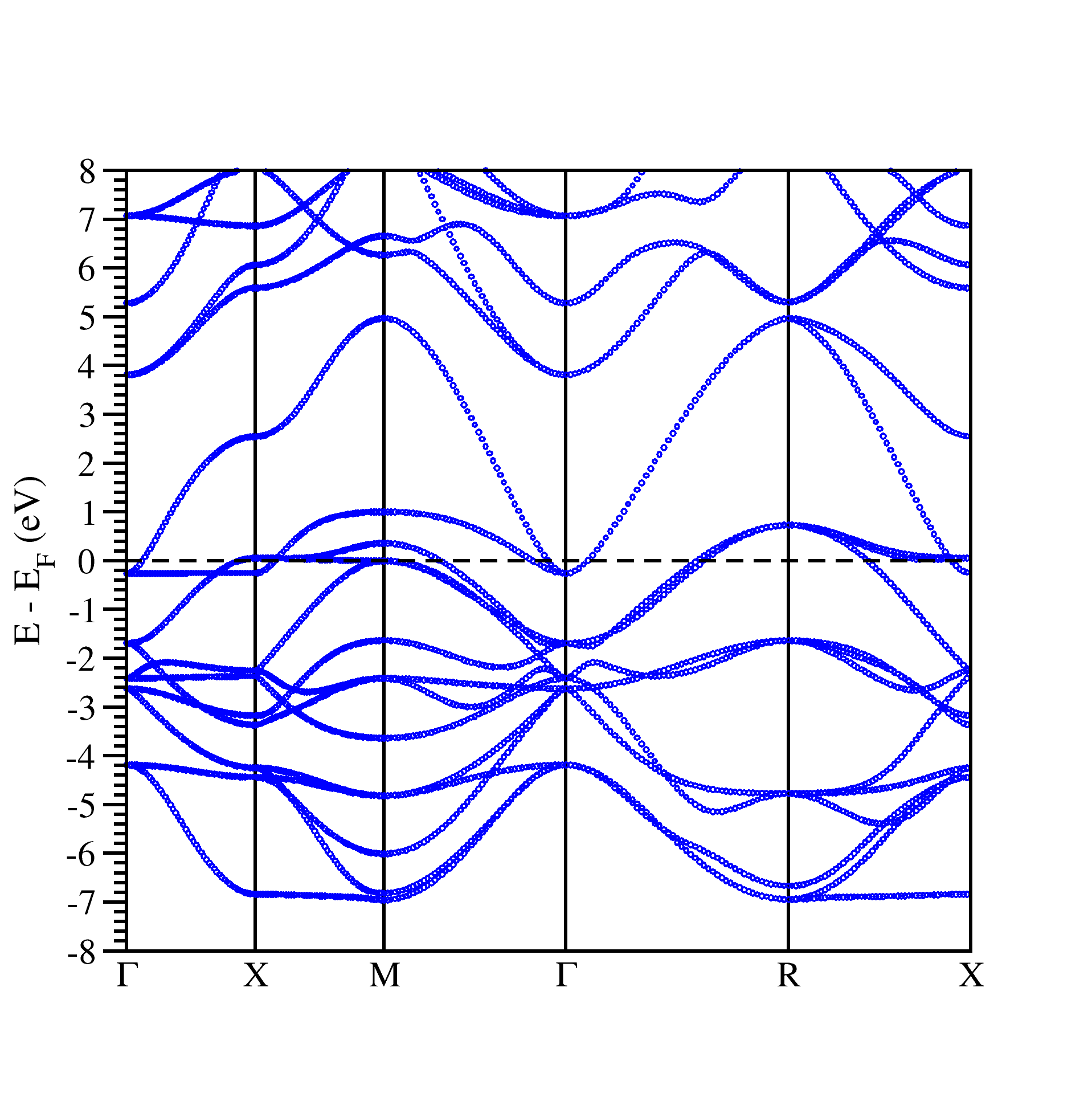}
\caption{(color online) Band-structure of cubic BaRuO$_{3}$ in its nonmagnetic phase.  Energies are
scaled to the Fermi-level (dotted line).}
\label{fig:fig1}
\end{figure}

\begin{figure}[t]
\centering
\includegraphics[angle=0,width=1.0\columnwidth]{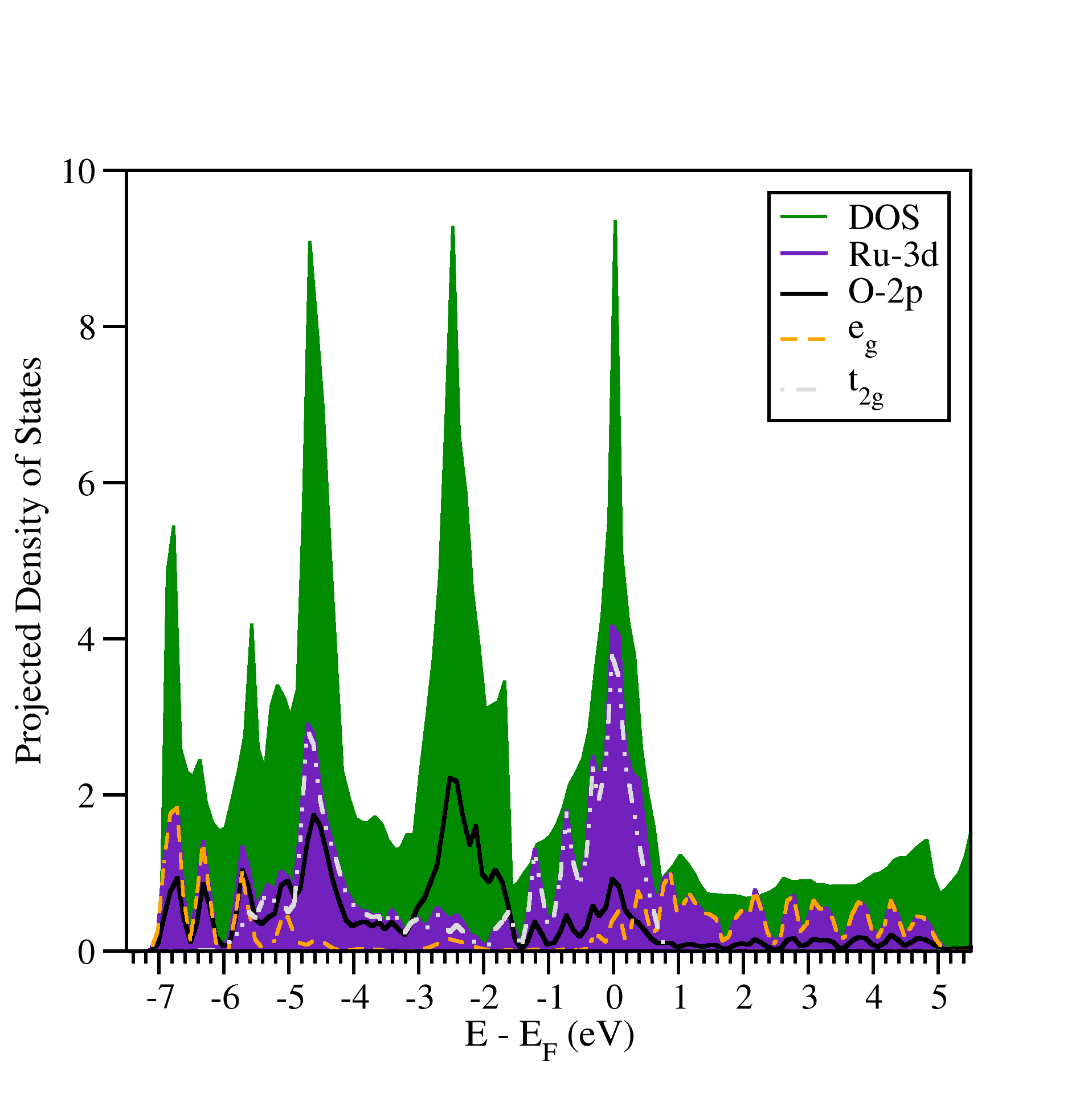}
\caption{(color online) Projected density of states (PDOS) of BaRuO$_{3}$. Green (shaded light gray),
violet (shaded dark gray), black (thick line), gray (dotted and dashed line) and orange (dashed
line) colors represents the DOS of whole system, Ru-atom, O-atom, Ru-t$_{2g}$ and Ru-e$_{g}$,
respectively.}
\label{fig:fig2}
\end{figure}

The electronic bandstructure, density of states (DOS) and projected DOS (pDOS) of BaRuO$_3$ in 
its non-magnetic (NM) phase are given in figures~\ref{fig:fig1} and ~\ref{fig:fig2}. 
The DFT results predict  BaRuO$_3$ to be a metal 
in its  non-magnetic phase with major contributions from the Ru-4d and O-2p orbitals 
across the Fermi-level. Hybridization between Ru-4d orbitals and O-2p orbitals spans from $\sim$
-8 eV below the Fermi level to $\sim$ 5 eV above the Fermi level. Bands above 5 eV are 
mainly composed of Ba-d orbitals and Ru-p orbitals. 

We find that, due to the octahedral environment of
the oxygen atoms surrounding the Ruthenium atoms, the
Ru-4d orbitals split into two sets, namely, t$_{2g}$ and e$_g$,
where t$_{2g}$ (e$_g$) orbitals contribution to the DOS is mainly
below (above) Fermi-level, supporting the low-spin t$_{2g}$ configuration of 
the nominal valence Ru$^{4+}$ (d$^4$). 

\begin{figure}[t]
\centering
\includegraphics[angle=0,width=1.05\columnwidth]{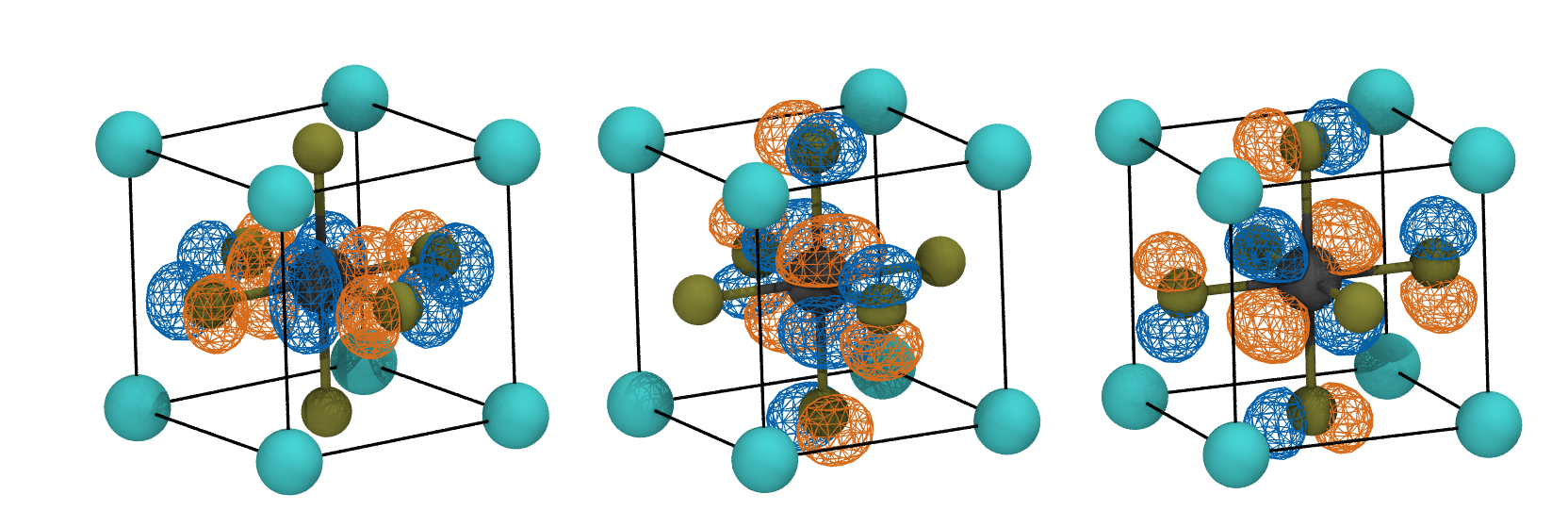}
\caption{(color online)  Orbital plots of maximally localized Wannier functions used to reproduce the
low energy subspace Hamiltonian.}
\label{fig:fig3}
\end{figure}

\begin{figure}[t]
\centering
\includegraphics[angle=0,width=1.05\columnwidth]{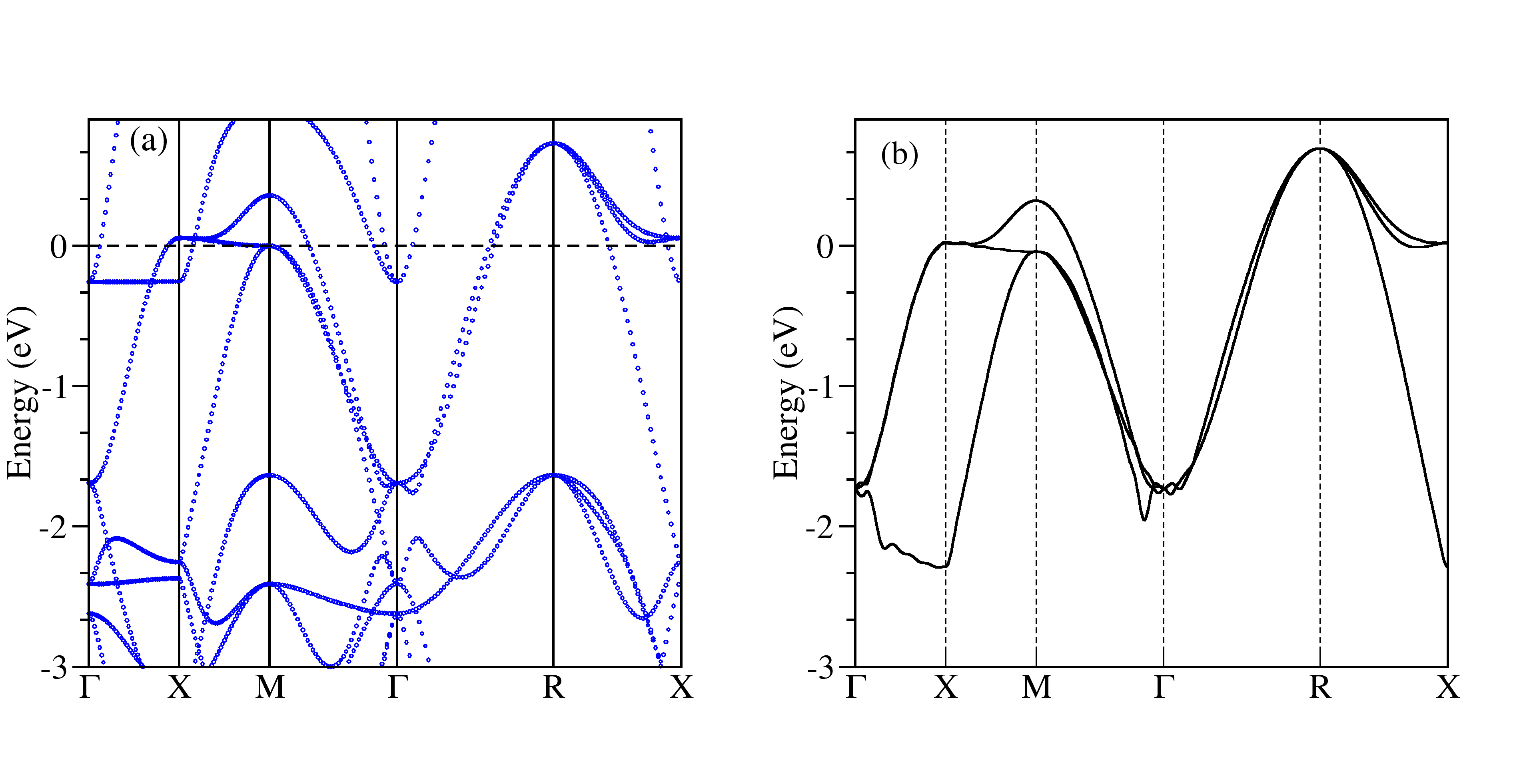}
\caption{(color online) Low energy subspace band-structure obtained from (a) Plane-wave basis and (b)
Wannier basis.}
\label{fig:fig4}
\end{figure}

From figure~\ref{fig:fig2}, we infer that the low energy subspace (-2.5 to 1 eV) which is 
relevant for the DMFT calculations is mainly composed of the Ru-t$_{2g}$ 
orbitals (with minor contributions of O-2p orbitals and Ru-e$_g$ orbitals) 
 have occupancy of $\sim$ 4 electrons.
Hence, to extract this low energy subspace Hamiltonian in an effective Wannier function basis, 
we have projected the Bloch-wave-functions obtained
from our DFT calculations onto the d$_{xz}$, d$_{yz}$, and d$_{xy}$
orbitals. The optimized Wannier functions calculated using
the MLWF method as implemented in Wannier90\cite{Mostofi2008685} code
are given in figure~\ref{fig:fig3} and the corresponding low energy
subspace band-structure calculated using these Wannier
functions are given in figure~\ref{fig:fig4}. Clearly, band-structures obtained 
from both the basis sets (Wannier, plane-wave) compare fairly well 
in the low energy subspace, validating the proper choice of our 
projections. Also, as shown in figure~\ref{fig:fig3}, the Wannier functions show the d$_{xz}$, d$_{yz}$, and
d$_{xy}$ orbital character and in addition have a substantial O-2p character 
due to their contributions near the Fermi-level.
The H({\bf{k}}) obtained in this Wannier basis is used for all the DMFT calculations, 
as the unperturbed or the `non-interacting' Hamiltonian.

\section{GGA+DMFT:}
In DMFT calculations we have introduced a local Coulomb interaction of density-density 
type between orbitals. The interaction part of the Hamiltonian is given 
in the second quantization notation by,
\begin{equation*}
H^{int}_{ii} = \sum^3_{i\alpha=1} U n_{i\alpha\uparrow} n_{i\alpha\downarrow} +
 \sum_{i\alpha\neq\beta} \sum_{\sigma\sigma'}
(V-J\delta_{\sigma\sigma'})n_{i\alpha\sigma} n_{i\alpha\sigma'}\,,
\end{equation*}
where $i$ represents the lattice site and $\alpha$, $\beta$ represent orbital 
indices. $U$ is the Coulomb repulsion between two electrons with opposite spin on the same 
orbital. We impose orbital rotational symmetry on the above Hamiltonian by setting $V = U-2J$, 
where $J$ is the Hund's coupling, which lowers the energy of a configuration with different 
orbitals ($\alpha$ $\neq$ $\beta$), and parallel spins $\sigma$ = $\sigma'$. We have solved 
the effective impurity problem within DMFT by using HY-CTQMC. In the literature, a range 
of $U$ and $J$ values have been used for 4d-Ru based TMOs. Indeed, determining these 
without ambiguity is not possible at present. In a recent work\cite{PhysRevB.91.195149}, 
using the constrained random 
phase approximation(cRPA) method, the $U$ value for ruthenates was 
found to be 2.3 eV. Thus, we choose $U_{Ru}$=2.3 eV. 
We fix the $J_{Ru}$ such that the theoretically calculated paramagnetic magnetic moment 
matches the corresponding experimentally measured value. Apart from this specific set of model 
parameters, we have investigated a range of ($U,J$) values in the neighbourhood of 
($U_{Ru}$,$J_{Ru}$) to ascertain the position of 3C-BaRuO$_3$ in the phase diagram. 
In the DMFT calculations, we find the chemical potential by fixing
the occupancy should be equal to 4 electrons per Ru, which is obtained from
threefold degenerate t$_{2g}$ bands in the Wannier basis or `non-interacting' Hamiltonian.
Now, we are going to discuss our results for single and two particle 
dynamics obtained from GGA+DMFT by using HY-CTQMC as an impurity solver.  
\subsection{Single Particle Dynamics:}

\begin{figure}[h!]
\centering
\includegraphics[angle=0,width=1.0\columnwidth]{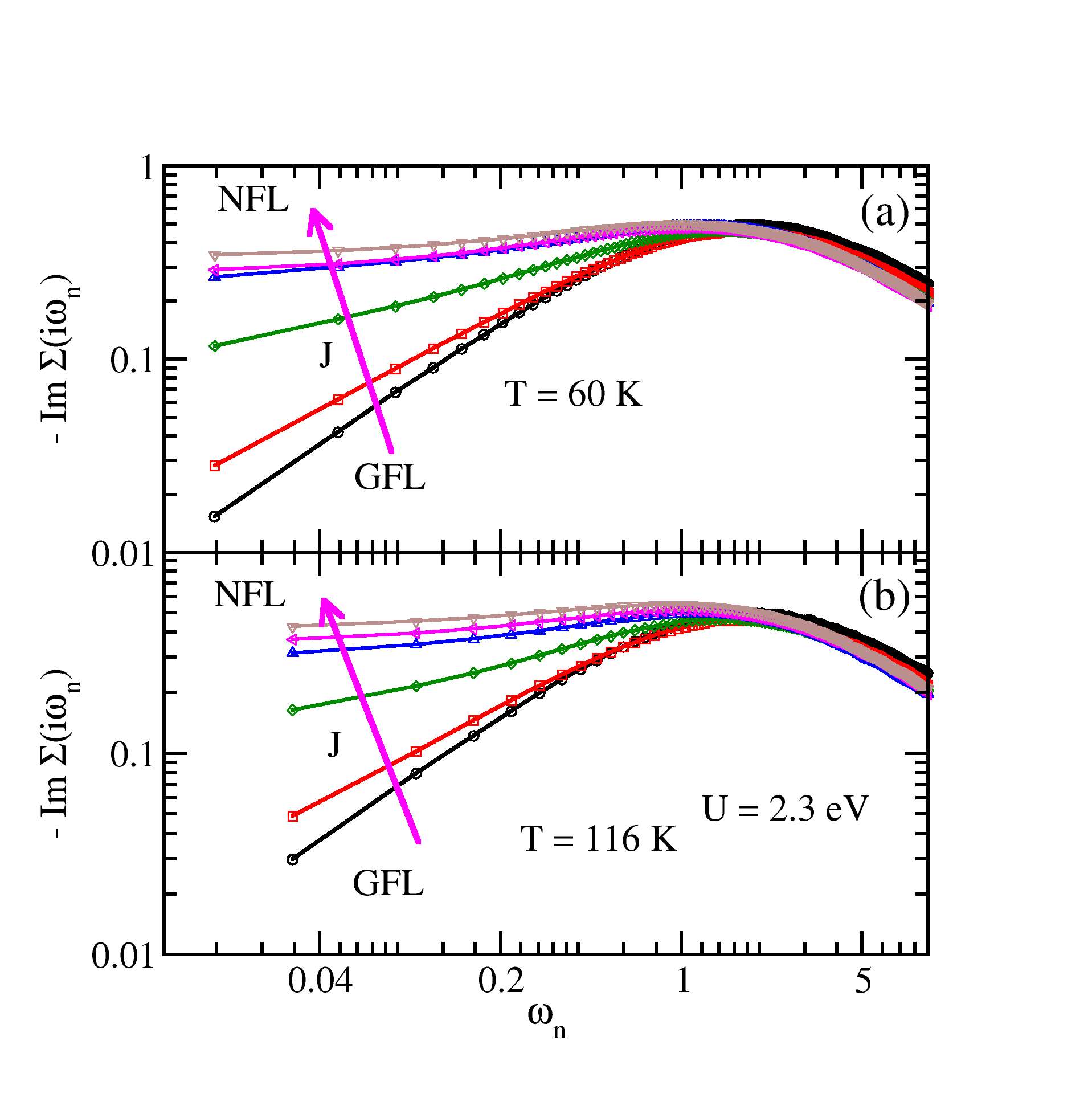}
\caption{(color online) Imaginary part of Matsubara self energy ($-{\rm Im} \Sigma (i\omega_n)$) 
for $U$ = 2.3 eV and different $J$ values for (a) $T$=60 K (b) $T$=116 K.}
\label{fig:fig5}
\end{figure}

To begin with, we focus on single particle dynamics that is mainly determined by 
the self-energy $\Sigma(i\omega_n$). Figure~\ref{fig:fig5}(a) shows
the imaginary part of Matsubara self-energy for $U = 2.3$
 eV and $T$ = 60 K for a range of 
$J$ values. For $J \lesssim 0.1$, the low-frequency behavior of self-energy has a generalized 
Fermi liquid (GFL) form i.e., $-{\rm Im}\Sigma(i\omega_n) \sim a\omega^{\alpha}_n$ 
where $0<\alpha\leq 1$. As we increase $J$, a deviation from
the power law is seen at low $\omega_n$ as the $-{\rm Im}\Sigma(i\omega_n)$ acquires a 
non-zero intercept. The latter is characteristic of non-Fermi liquid behaviour, 
where the imaginary part of self-energy has a finite value as $\omega_n \rightarrow 
0$. Thus 
as a function of increasing $J$, the single particle dynamics exhibits a crossover from
GFL to NFL that is 
driven by Hund's exchange\cite{PhysRevLett.101.166405}. 
The crossover is found to persist at a  higher 
temperature $T$= 116 K and is shown in figure~\ref{fig:fig5}(b).

\begin{figure}[t]
\centering
\includegraphics[angle=0,width=1.0\columnwidth]{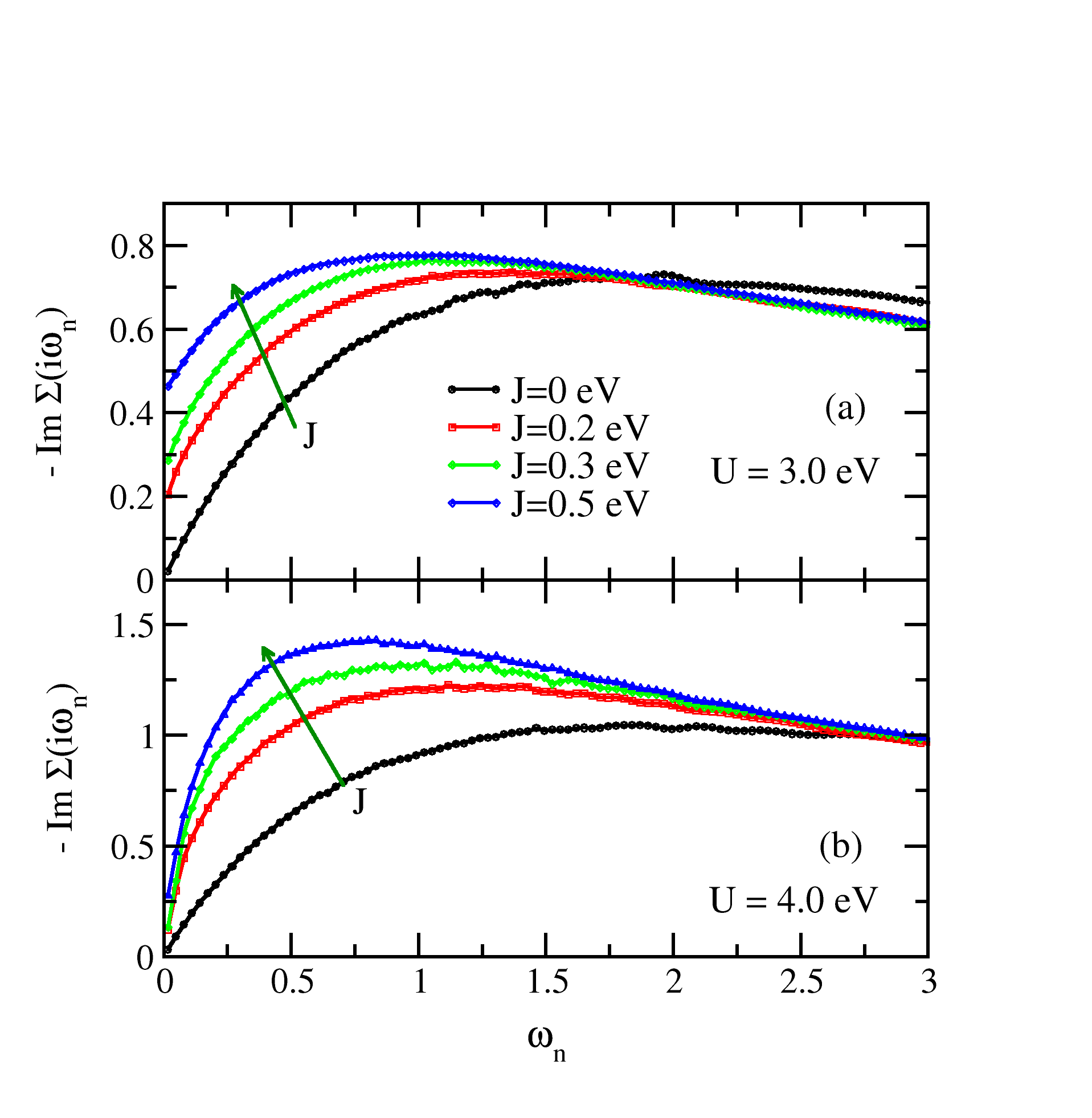}
\caption{(color online) Imaginary part of Self energy for $T$=60 K and 
different $J$ values (mentioned in legends) with (a) $U$ = 3 eV,  and (b) $U$ = 4 eV.}.
\label{fig:fig11}
\end{figure}

A natural question arises about the choice of the $U=2.3$eV
for 3C-BaRuO$_3$. Does this crossover from GFL to NFL survive with respect to variations in $U$?
The imaginary part of self-energy for 
$U=3$ and 4eV computed at a temperature, $T$=60 K 
is shown in figure~\ref{fig:fig11}. Clearly, for $U=2.3$
and 3 eV, the intercept of the imaginary part of the self-energy is 
finite for $J\gtrsim 0.2$ (from figure~\ref{fig:fig5} and the 
top panel of figure~\ref{fig:fig11}), while for $U=4$ eV,
a GFL form of  $-{\rm Im}\Sigma(i\omega_n)$ is obtained
for $0\leq J \leq 0.5$ eV. This implies that the NFL behaviour for 
higher values of $U (\gtrsim 4)$ eV, if at all occurs, must be for $J > 0.5$eV. 
Hence, we conclude that the $U_{Ru}=2.3$eV, corresponding to 3C-BaRuO$_3$ is
somewhat special, since it places this material in a
 crossover region for
{\em physically reasonable} values of the Hund's exchange.

It is known from recent works on ruthenates that the NFL
behaviour seen in the single-particle dynamics is characteristic of a 
finite temperature spin-frozen phase
which crosses over to a Fermi liquid ground state at lower temperatures. 
This incoherent spin-frozen state\cite{PhysRevLett.106.096401} is characterised by 
finite intercepts in the imaginary part of self-energy and fluctuating local moments (through
susceptibility). In order to understand the 
crossover phase in a better way, we 
\begin{figure}[h!]
\centering
\includegraphics[angle=0,width=1.0\columnwidth]{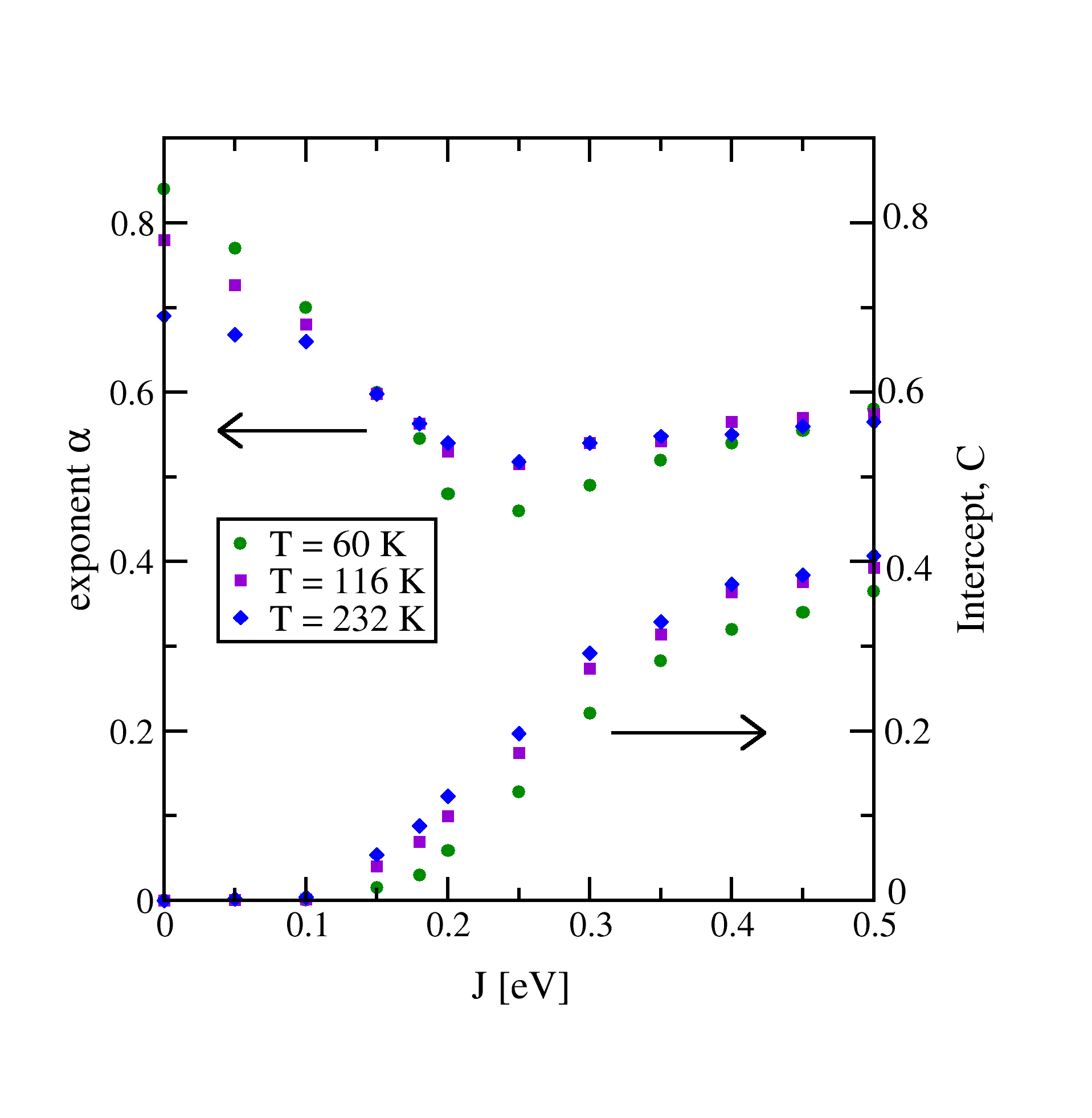}
\caption{(color online) Exponent $\alpha$ (left) and intercept $C$ (right) obtained
 by fitting the data 
to -Im $\Sigma(i\omega_n)$ = $C$ + $A$ $|\omega_n|^{\alpha}$ at different $J$ values, 
$U$ = 2.3 eV and 
$T$ = 60 K, 116K and 232K.}
\label{fig:fig8}
\end{figure}
carry out a quantitative analysis of
the imaginary part of the self-energy for many more $J$ values in the 
same range as considered in figure~\ref{fig:fig5}. The imaginary
part of self-energy at low Matsubara frequencies 
is fit to the form\cite{PhysRevLett.101.166405}
\begin{equation}
 -{\rm Im}\Sigma(i\omega_n) \stackrel{\omega_n\rightarrow 0}{\rightarrow} C +A |\omega_n|^{\alpha}\,,
 \label{eq:pfit}
 \end{equation}
  and  
figure~\ref{fig:fig8} shows the  exponent $\alpha$ (circles) and 
intercept $C$ (squares) as a function of
$J$ at various temperatures from 60K to 230K,
 for $U$ =2.30 eV. The exponent $\alpha$ 
 initially decreases with increasing $J$, goes through a minimum value of 
0.5 at a $J\sim 0.25$eV and increases
 gradually for higher $J$. Such behaviour has been
 found previously by Werner\cite{PhysRevLett.101.166405} et. al., in  
a three orbital Hubbard model with fully rotationally invariant interactions for 
fixed filling ($n$=2.0) and Hund's exchange, but varying the 
$U$ value. 
The intercept $C$ remains zero for $J\lesssim$ 0.15 eV and above that it has
a finite value which increases with $J$. Thus we identify a crossover 
Hund's exchange $J_0 = 0.15$ eV such that for
$J < J_0$ the GFL phase exists, while for $J > J_0$ the
 crossover NFL phase is found for $\gtrsim 60$K, where frozen 
moments are expected to scatter the conduction electrons.
 It is interesting to note that 
the exponent $\alpha$ in the GFL or in the NFL region is not equal to 1.  In the GFL phase,
the exponent must approach 1 with decreasing temperature, and indeed, it does, as seen
in figure~\ref{fig:fig8} for $J< J_0$. Curiously, the exponent  hardly changes with either temperature
or $J$ in the spin-frozen phase even until 60K. For 3C-BaRuO3, a ferromagnetic transition occurs at
$T_c=60K$. Thus, it is likely that the spin-frozen phase is a precursor of the FM phase, and
the local moments condense into a magnetically ordered state for $T< 60K$. 
 We have repeated the above analysis for $U=3$eV and find that the 
crossover $J_0\sim 0.15$eV is the same as that for
 $U=2.3$ eV within numerical tolerance.  Even the intercept depends very weakly on temperature, thus,
 the spin-frozen phase appears to be almost temperature independent. This implies that
 the NFL behaviour should manifest in transport and thermodynamic quantities over a wide range
 from about 60K to at least 230K.

\begin{figure}[h!]
\centering
\includegraphics[angle=0,width=1.0\columnwidth]{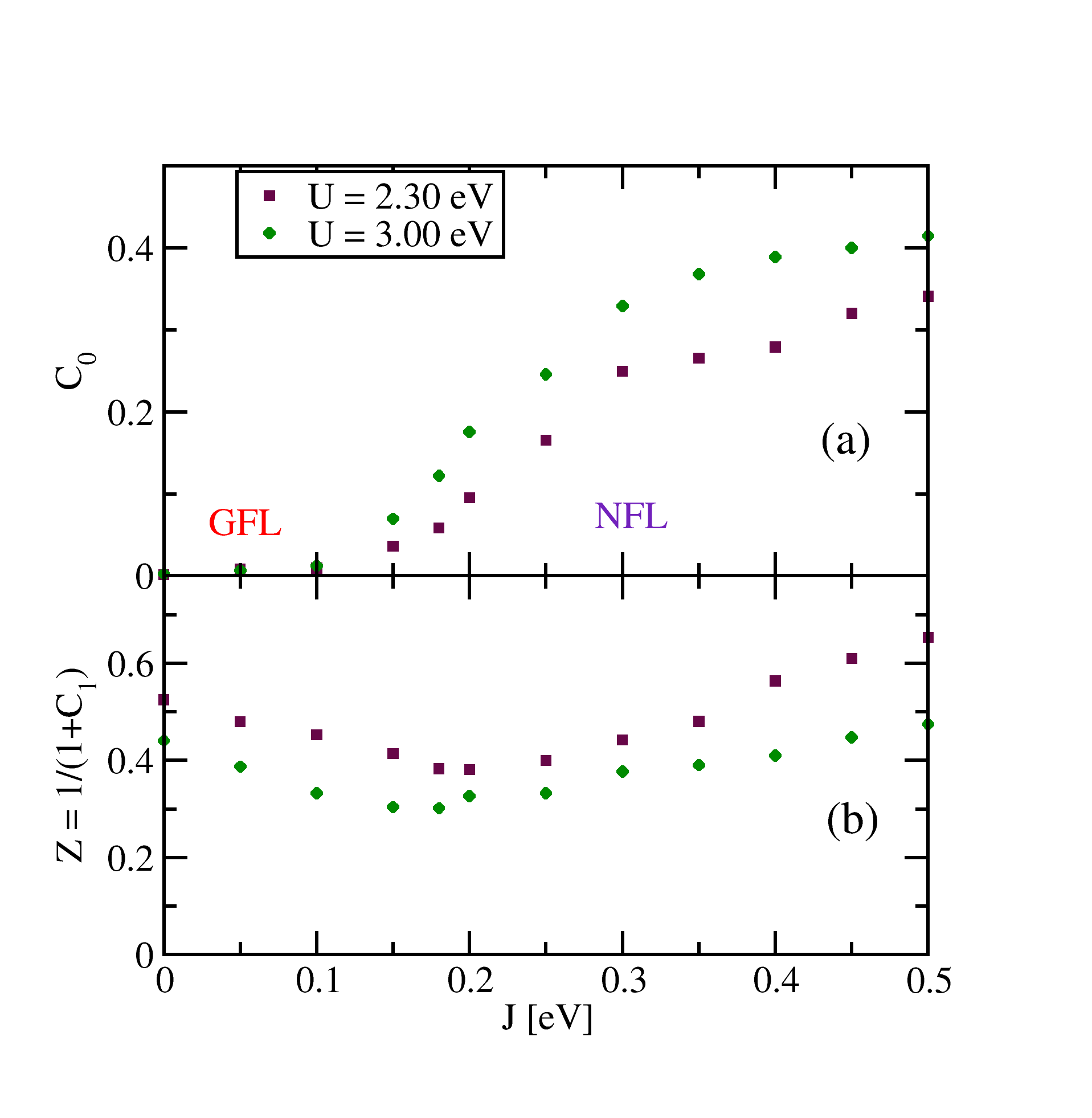}
\caption{(color online) Imaginary part of self energy (-Im $\Sigma$(i$\omega_n$)) fitted to 
4th order polynomial: (a) zeroth order coefficient, $C_0$ (b) $Z=1/(1+C_1)$, 
where $C_1$ is the linear coefficient, for 
different $J$ values, $U$=2.3 and 3.0 eV and $T$ = 60 K.}
\label{fig:fig10}
\end{figure}

The crossover function, given in
equation~\ref{eq:pfit} does not have a microscopic basis, and has been 
used purely as a fitting function. Since the latter is not unique, 
the identification of $J_0$ must be verified through an alternative fit. 
Hence, we have used a fourth order polynomial also to fit $-{\rm Im}\Sigma(i\omega_n)$  and confirm
the robustness of $J_0$. 
The intercept $C_0$ shown in the top panel of figure
~\ref{fig:fig10} does become non-zero only for $J\gtrsim J_0$.
Thus, the identification of $J_0$ remains robust.
For a Fermi liquid, the linear coefficient of the self-energy, $C_1$ is 
related to the quasiparticle weight, $Z$ by
$C_1=-(1-1/Z)$ at $T=0$. Although $C_1$ does not have the same 
interpretation at finite temperature, a qualitative 
picture may be obtained by examining the dependence 
of $Z=1/(1+C_1)$. The lower panel of 
figure~\ref{fig:fig10} shows that the $Z$ 
decreases throughout the GFL phase. Although the
$Z$ lacks any interpretation in the NFL phase ($J> J_0$),
a finite $Z$ is, nevertheless, obtained which behaves in a similar 
way as the exponent of the power law fit (figure~\ref{fig:fig8}).

\subsection{Two Particle Dynamics:}

\begin{figure}[t]
\centering
\includegraphics[angle=0,width=1.0\columnwidth]{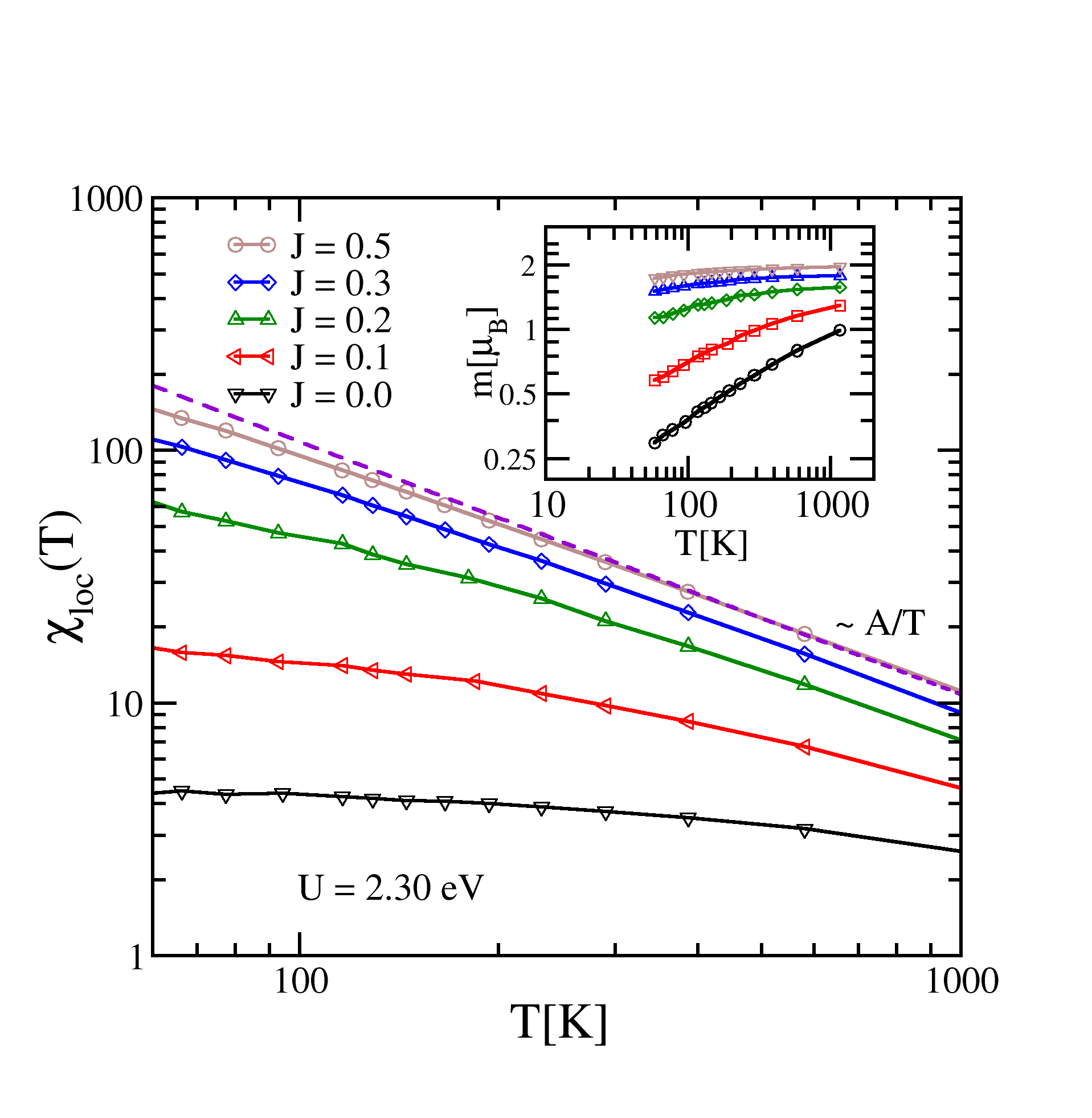}
\caption{(online) Local static spin susceptibility as a function of temperature 
for different $J$ values and $U$ = 2.30 eV. The dashed curve represents a $1/T$ fit at high temperatures. 
Inset: The screened magnetic moment as a function of temperature. }
\label{fig:fig7}
\end{figure}

The effect of temperature on spin correlations may be gauged through
the local static spin susceptibility, given by $\chi_{loc}(T)$ = 
$\int^{\beta}_0 d\tau \chi_{zz}(\tau)$. Figure~\ref{fig:fig7}, shows 
$\chi_{loc}(T)$ as a function of temperature for a range of $J$ values. 
For $J\lesssim 0.1$, $\chi_{loc}(T)$ is very weakly dependent of temperature 
over the entire range shown, which is 
characteristic of Pauli-paramagnetic behavior and hence corresponds to a GFL behaviour. 
For larger $J$ values, we observe local moment behavior ($\chi_{loc}(T)\sim \frac{1}{T}$) behaviour
at lower temperatures as well (see dashed line fit in the main panel). Thus with 
increasing $J$, $\chi_{loc}$ also crosses  over to local moment region from GFL regime. 
We will see later that the temperature dependence of susceptibility allows to
identify the value of Hund's exchange coupling appropriate for 3C-BaRuO$_3$. 
The inset shows the screened magnetic moment
as a function of temperature computed through\cite{PhysRevB.21.1003,PhysRevLett.104.197002} 
$m$ = $\sqrt{T\chi(T)}$. In the GFL phase
($J<J_0$), the magnetic moment is seen to decrease monotonically 
with decreasing temperature indicating an absence of local moments at $T=0$. While for $J>J_0$, the
magnetic moment {\em appears} to saturate as $T\rightarrow 0$
indicating fluctuating incoherent local-moments in the 
spin-frozen phase.

In most of the 4d Ru-based TM oxides, most theoretical studies are restricted to single-particle
spectral functions and static susceptibilities\cite{PhysRevLett.106.096401,1367-2630-11-2-025021}.
There are only a few studies on two particle spectral functions including vertex
corrections\cite{PhysRevB.86.064411}, and even those are limited to fixed $U$ and $J$ values.
However, there are no studies available for the behavior of two particle spectral functions
(including vertex corrections) across the GFL to NFL crossover.
   
\begin{figure}[t]
\centering
\includegraphics[angle=0,width=1.0\columnwidth]{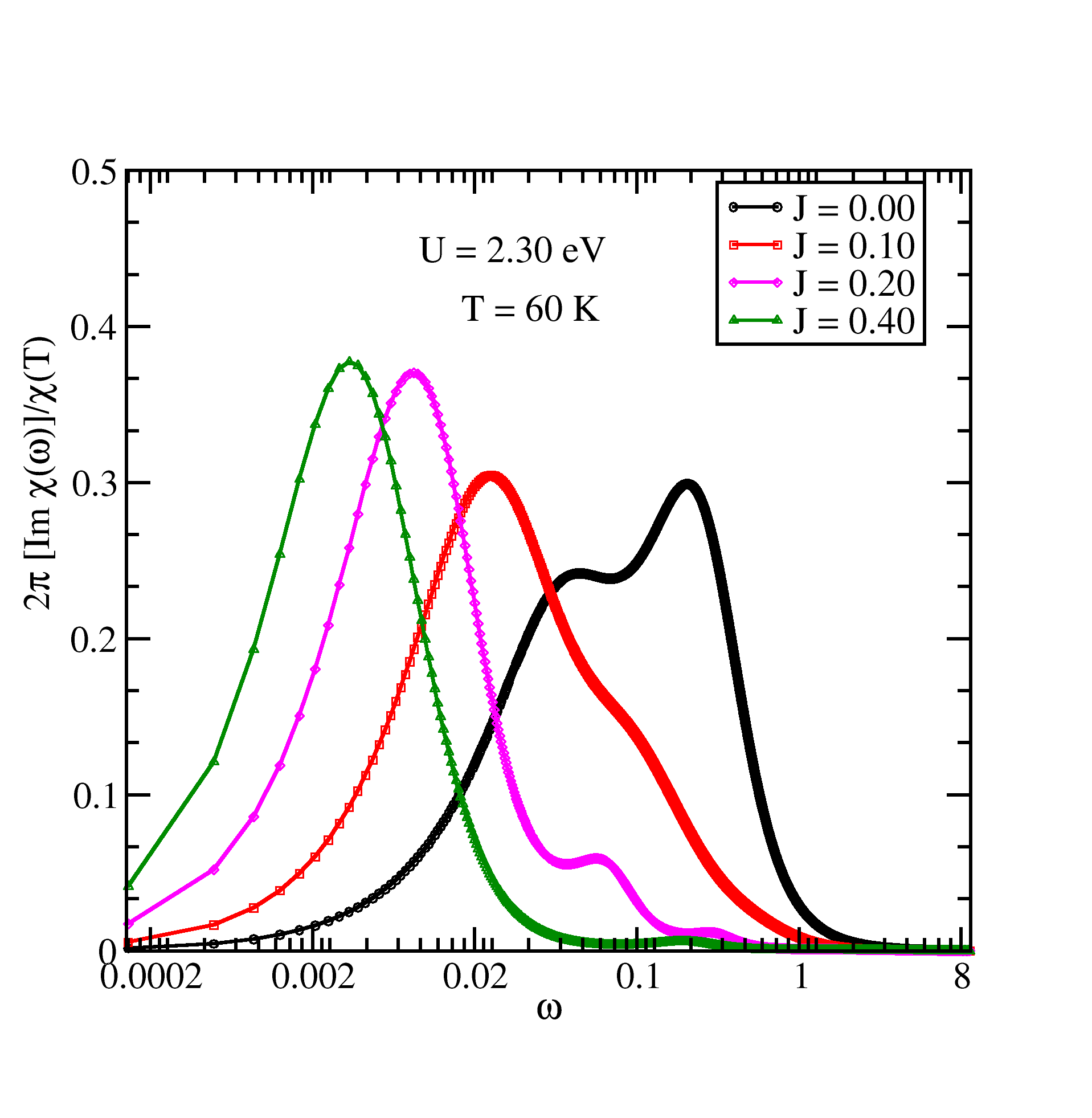}
\caption{(color online) Imaginary part of dynamical spin susceptibility on 
real frequency axis obtained from maximum entropy method for various $J$ values, $U$ = 2.3 eV
 and $T$ = 60 K.}
\label{fig:fig12}
\end{figure}

  We have calculated the dynamical spin susceptibility $\chi(\omega,T)$ on the real frequency axis
by using maximum entropy method\cite{MAX,jarrell,dasari1}. In figure~\ref{fig:fig12}, we show the
imaginary part of $\chi(\omega,T)$ for various $J$ values at $U$ = 2.30 eV and $T$=60 K. A large
scale spectral weight transfer to the infrared occurs upon increasing $J$ of $\chi(\omega,T)$.
Concomitantly, the half-width at half maximum also decreases.  The peak in $\chi(\omega,T)$
represents the characteristic energy scale of the system\cite{PhysRevB.44.5347,PhysRevB.86.064411},
below which a Fermi liquid should emerge. The dramatic red shift of the peak with increasing $J$
implies a strong suppression of the coherent scale\cite{PhysRevB.79.115136,PhysRevB.44.5347,dasari}.
Thus with increasing $J$, the energy scale for crossover from a low temperature Fermi liquid ground
state to a high temperature incoherent phase decreases sharply. Since the only other scale (apart
from the coherence scale) are the non-universal scales such as $J$ or the bandwidth or $U$, the
incoherent crossover phase should exist from very low temperatures to quite high temperatures. This
explains the wide temperature range over which an incoherent spin-frozen phase, and the
corresponding non-Fermi liquid behaviour is found, e.g in the
resistivity\cite{PhysRevB.53.4393,Georges,Jin20052008}.


\subsection{Identification of $J$ for 3C-BaRuO$_3$}

Now we turn to an identification of model parameters appropriate for 3C-BaRuO$_3$ in the ($U,J$) plane. 
As mentioned earlier, we have chosen $U_{Ru}$=2.3 eV for 3C-BaRuO$_3$ which has been
 obtained through cRPA for its closely related cousins in the ruthenate 
family\cite{PhysRevB.91.195149,PhysRevLett.106.096401,PhysRevB.75.035122}. 
The $J_{Ru}$ is obtained by comparing the theoretically computed, temperature dependent,
 static susceptibility (from figure~\ref{fig:fig7}) with that of the experiment\cite{Jin20052008}. From 
experiments, it is known that the saturated magnetic moment at 5K (in the ferromagnetic
state) is $0.8 \mu_B$/Ru, while the high temperature paramagnetic moment is $2.6 \mu_B$/Ru.
Since our theory is valid only in the non-magnetic phase, we choose the latter for theoretical comparison.
One more issue in the theory is the use of Ising-type or density-density type Hund's coupling,
which results in a $S=1$ state corresponding to an ideal magnetic moment of $2\mu_B$/Ru rather than
$2.8 \mu_B$ as would be expected for a true $S=1$ state with a rotationally invariant $J$ term.
Thus, the high temperature moment that we would be comparing to is $(2.6/2.8)\times 2 = 1.86 \mu_B$/Ru.
We see from the inset of figure ~\ref{fig:fig7} that such a moment is obtained for $J\sim 0.5$ eV.
Hence we identify $J_{Ru}\sim 0.5eV$. We note that the experimentally measured $\chi_{loc}^{-1}(T)$ 
is linear at high temperature, and deviates from linearity\cite{Jin20052008} at 
$T\lesssim 150K$. Again, such deviation from the high temperature $1/T$ form 
in theoretical calculations is seen for $J\sim 0.5$ at $T\lesssim 150K$ (in the main panel
of figure~\ref{fig:fig7}), thus lending support  to the identification of $J_{Ru}\sim 0.5$ eV 
from the magnetic moment. We have checked that the deviations from linearity occur 
at much higher temperatures ($\gtrsim 300K$) for $J=0.3$ and 0.4eV,
hence the error bar on $J_{Ru}$ should be less than 0.1eV.

The value of Hund's coupling $J_{Ru} \sim 0.5$ eV places 3C-BaRuO$_3$ deep in the
incoherent spin-frozen phase for $T\gtrsim 60K$, and thus could explain the 
transition into a magnetically ordered state at $T\lesssim 60K$.
The experimentally observed non-Fermi liquid behavior in $\rho(T)$ could originate 
from an anomalous self-energy. Indeed as figure~\ref{fig:fig7} shows, 
the self-energy at the chemical potential has a finite and almost 
temperature-independent imaginary part. In addition to the static part, 
it would be interesting to see if the dynamics also contributes to the NFL behaviour.
\begin{figure}[t]
\centering
\includegraphics[angle=0,width=1.1\columnwidth]{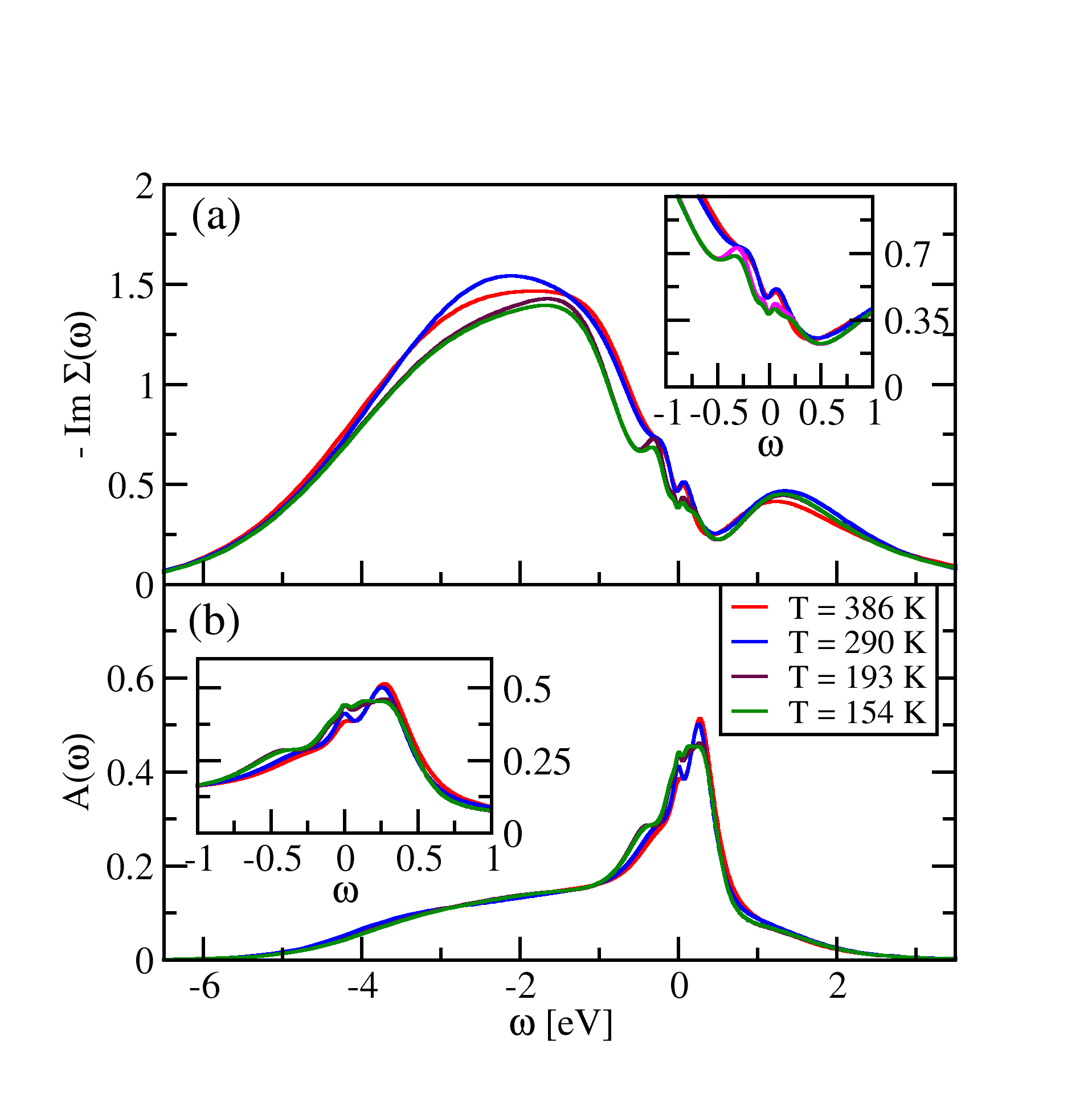}
\caption{(color online) (a) Imaginary part of self-energy (b) single particle
spectral function on real frequency axis obtained from maximum entropy method
for different temperatures and $U$ = 2.3 eV, $J$ = 0.5 eV.}
\label{fig:fig13}
\end{figure}
 Hence, we compute the real frequency self-energy through analytic 
continuation of the Matsubara $\Sigma(i\omega_n)$ and display
${-\rm Im}\Sigma(\omega)$ (top panel) and the corresponding 
$k$-integrated spectrum, $A(\omega)=-{\rm Im}G(\omega)$ (bottom panel) 
for various temperatures in figure~\ref{fig:fig13}, where the local Green's function is given by 
\begin{align}
\mathbf{G}(\omega) & =\sum_k \mathbf{G}(k,\omega)\nonumber \\
& = \sum_k \frac{1}{(\omega^+ +\mu)\mathbf{I}-\mathbf{\mathcal{H}_{GGA}}(k)-\mathbf{\Sigma}(\omega)}\,.
\label{eq:gkw}
\end{align}
Note that, within DMFT, the $k$-dependence arises
purely through the dispersion embedded in $\mathcal{H}_{GGA}(k)$.
If the low energy excitations are
Fermi-liquid like, then we should expect Im $\Sigma(0)\propto -T^2$.
However the almost temperature-independent and finite value of 
Im $\Sigma(0)$ shown previously in figure~\ref{fig:fig8} and also 
seen in figure~\ref{fig:fig13}(a) signifies that low energy excitations are
NFL in nature, and temperature does not have much effect on the value of Im $\Sigma(0)$ 
in the spin-frozen phase. A very interesting insight into the dynamics of the spin-frozen phase comes
from the low frequency form of the self-energy. The inset zooms in onto the low frequency 
part of ${-\rm Im}\Sigma(\omega)$, which is seen to have a form $\sim C+A\omega^2$ that 
is usually found in disordered Fermi liquids\cite{PhysRevB.88.195120}. Such a form is consistent 
with the scenario of incoherent and fluctuating local moments in the spin-frozen phase. 
The single-particle spectral function  $A(\omega)$ shown in figure~\ref{fig:fig13}(b) 
has an overall lineshape very similar to that of
SrRuO$_3$\cite{PhysRevB.91.205116} and Sr$_2$RuO$_4$\cite{Georges}. 
A metallic nature is indicated by a finite weight at the Fermi level. A closer look
at the temperature dependence at low frequencies shows the emergence of structures 
that presumably correspond to transitions between the various multiplets of the
atomic limit. However, a far more detailed study, varying
$U$ and $J$, is required for a precise identification of 
the origin of these features. Since
the procedure of analytic continuation using the maximum
entropy method requires immense computational resources,
especially in the spin-frozen phase, we have not attempted
to carry out such a  study in the present work. 

\begin{figure}[t]
\centering
\includegraphics[angle=0,width=1.0\columnwidth]{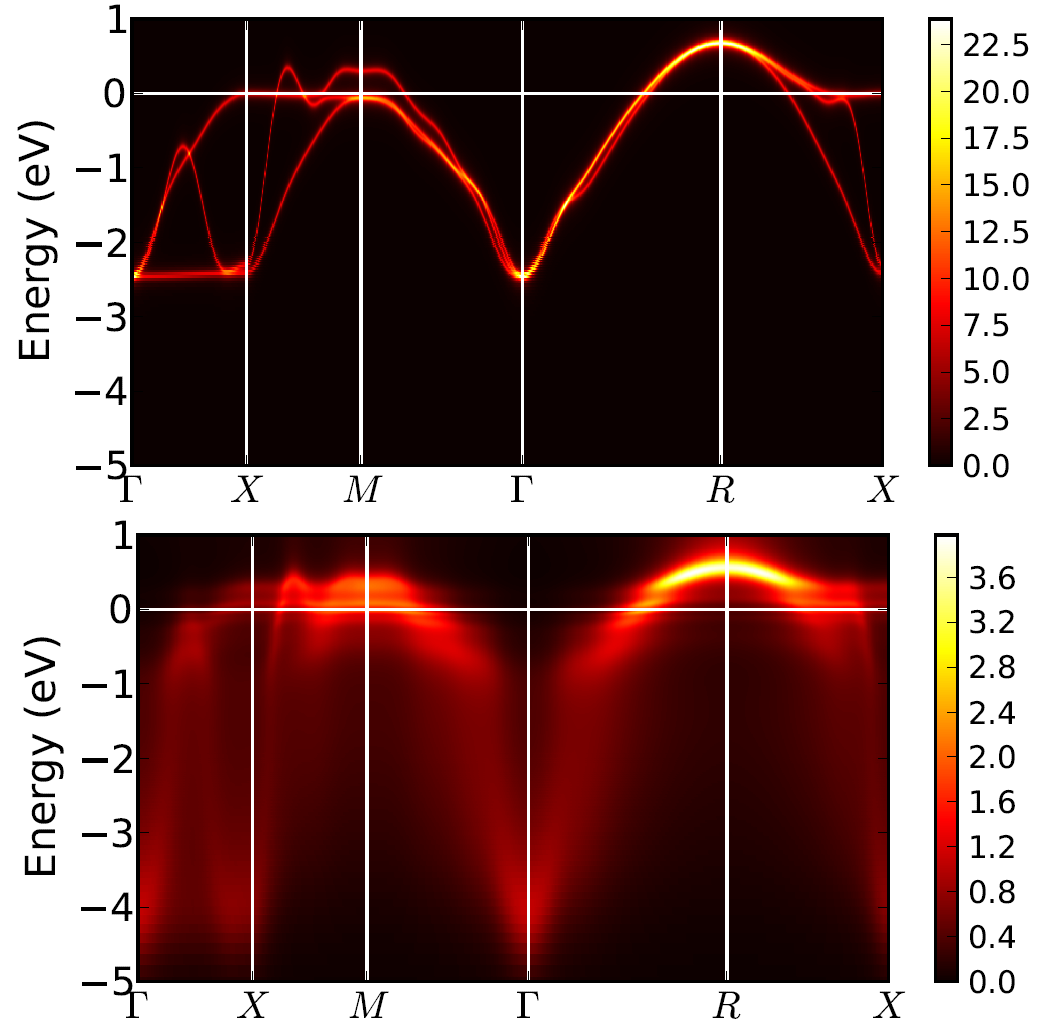}
\caption{(color online) Intensity map of the spectral function A(k,$\omega$) 
obtained from DFT (top panel) and DFT+DMFT (lower panel) for $U$ = 2.3 eV, $J$ = 0.5 eV 
at T=154 K plotted along high symmetry directions in the irreducible Brillouin zone.}
\label{fig:fig14}
\end{figure}

Experiments can probe single-particle dynamics in the spin-frozen phase through e.g, 
angle-resolved photoemission spectroscopy (ARPES). Theoretically we can predict 
the ARPES lineshape through a calculation of the $k$-resolved spectral function is
given by $A(k,\omega)= -{\rm Im} G(k,\omega)/\pi$.
In figure~\ref{fig:fig14} we have plotted the intensity map of the momentum-resolved spectral 
function $A(k,\omega)$ of 3C-BaRuO$_3$ obtained from DFT (top panel, obtained by simply ignoring the self-energy in equation\ref{eq:gkw}) 
and then compared with the results of DFT+DMFT (lower panel) at $T$ = 154 K 
for $U_{Ru}$= 2.3 eV and $J_{Ru}$ = 0.5 eV. In case of DFT, the quasi-particle 
bands have a minimum at $\Gamma$ point (-2.3 eV) and maximum at $R$
point (0.8 eV). When we turn on interactions (in the case of DFT+DMFT), 
the first striking feature that emerges is that there are no quasiparticle bands. 
In other words, each of the bands obtained within DFT acquires a finite width when
interactions are introduced, but there does exist a resemblance of the quasiparticle bands in the
spectral function map. The `fat bands' are simply a result of the finite scattering rate 
arising from the imaginary part of the self-energy (figure~\ref{fig:fig13}). Furthermore, although the
bands in DFT as well as DFT+DMFT have a minimum and maximum exactly at the same 
high symmetry points, the values of corresponding energies renormalize to -4.8 and 1.0 eV 
respectively in the latter. The bands below -1.0 eV 
(incoherent regime) are much more broadened\cite{PhysRevLett.106.096401,PhysRevB.91.205116} 
in comparison with the those closer to the Fermi-level, which again is a manifestation of 
the peak in the imaginary part of the self-energy around -2 eV. 

For the values of $U_{Ru}$=2.3 eV and $J_{Ru} \sim 0.5$ eV, we obtain a relatively modest 
effective mass $\frac{m^*}{m_{GGA}}$ of $1.56$ at $T=60K$. A definitive comment about the effective
mass in the ground state cannot be made with the preceding estimate at finite temperature, since
the quasiparticle weight has a proper meaning\cite{PhysRevB.92.075136} only below the Fermi liquid coherence scale, which
is strongly suppressed for $J=0.5$ (as compared to $J=0$)
 as seen from the dynamical susceptibility results (from figure
~\ref{fig:fig12}). Thus, unless extremely low temperature calculations are carried out,
a proper estimate of $m^*$ is not possible. Nevertheless, the strong suppression of the Fermi
liquid scale suggests that 3C-BaRuO3 could be very strongly correlated.
Here, we would like to comment on the value of $U$(= 4.0 eV) and $J$(=0.6 eV) chosen in a 
previous work \cite{PhysRevB.87.165139} on 3C-BaRuO$_3$ within the dpp model. 
They obtained the interaction parameters from  a ``local spin density approximation constraint"
technique. For those parameters, a recent study of one of the 4d Ruthenium compounds\cite{PhysRevB.91.195149}
within a five d-band model finds that correlations are induced due to the proximity of a 
Mott insulating state, which concurs with  our results for a three d-band 
model (from the lower panel of figure~\ref{fig:fig2}). However, the proximity of a 
Mott insulating state does not violate adiabatic continuity and hence as shown above, 
the choice of $(U,J)=(4.0,0.6)$ eV would not explain several anomalous features of 
3C-BaRuO$_3$ including the wide $1/T$ behaviour of $\chi_{loc}(T)$, or the
NFL behaviour of resistivity. These and the transition to a ferromagnetically 
ordered state at low temperature are naturally explained by the presence of a spin-frozen phase
as found for $U_{Ru}=2.3$eV and $J_{Ru}\sim 0.5$ eV.

\section{Conclusions}

We have studied the 3C-BaRuO$_3$ in the non-magnetic phase by using GGA+DMFT (HY-CTQMC). 
In the dynamical correlation functions and static spin 
susceptibility, we observed a crossover from GFL to NFL driven by the Hund's exchange 
$J$ and a fitting of the self-energy to a power law function 
function ($\omega^{\alpha}_n$) determined the cross-over boundary 
i.e., $J_0$ = 0.15 eV. The local, on-site Coulomb repulsion, $U_{Ru}$ = 2.3 eV, was
chosen to be the same as that found through constrained random phase approximation calculations
for the closely related SrRuO$_3$. We determine the Hund’s exchange, $J_{Ru}$, appropriate for
3C-BaRuO$_3$ such that the computed high temperature paramagnetic moment matches the experimentally 
found value, and thus we find that $J_{Ru}$ $\sim$ 0.5 eV. Non-magnetic calculations with these parameters 
($U_{Ru}$, $J_{Ru}$) for single-particle dynamics and static spin susceptibility show that 
cubic-BaRuO3 is in a spin-frozen state at temperatures above the
ferromagnetic transition point. Future calculations incorporating symmetry broken states should reveal the 
causal relation between the high temperature spin-frozen phase and the dynamics
in the low temperature ferromagnetic phase. 
\section{Acknowledgments}

We thank CSIR and DST (India) for research funding.
This work is partly supported by NSF DMR-1237565 (NSV and JM) and NSF EPSCoR Cooperative Agreement
EPS-1003897 with additional support from the Louisiana Board of Regents (ND and MJ). ND thank Swapan K. Pati for helpful discussions. Our simulations used an open source implementation~\cite{Hafer} of the hybridization expansion
continuous-time quantum Monte Carlo algorithm~\cite{Comanac} and the ALPS~\cite{Bauer}
libraries. The computational resources for CTQMC simulations are provided by
the Louisiana Optical Network Initiative (LONI) and HPC@LSU.


\bibliographystyle{apsrev4-1}
\bibliography{apssamp}
\end{document}